\documentclass{article}
\usepackage{graphicx}
\usepackage{amssymb}
\begin{document}

\title{Computational  Dating  for the Nuzi Cuneiform  Archive: The  Least 
Squares Constrained by Family Trees and  Synchronisms}


\author{Sumie Ueda \thanks{
ueda@ism.ac.jp,  
Institute of Statistical Mathematics, 10-3 Midori-cho, Tachikawa Tokyo 190-8562 Japan } 
Takashi Tsuchiya \thanks{tsuchiya@grips.ac.jp, 
National Graduate Institute for Policy Studies, 7-22-1 Roppongi 
 Minatoku Tokyo 106-8677 Japan}
 Yoshiaki Itoh \thanks{
itoh@ism.ac.jp, 
Institute of Statistical Mathematics and Graduate University for Advanced Studies, 10-3 Midori-cho Tachikawa Tokyo 190-8562 Japan} } 
\maketitle

\begin{abstract}
We introduce a computational method of dating for an archive in ancient Mesopotamia. We use the name index  Nuzi Personal Names (NPN)  published in 1943.    We made an electronic version of NPN  and added the kinships of the two powerful families to NPN to reflect the Nuzi studies after 1943.  
Nuzi is a town from the 15th - 14th century B.C.E.
for a period of some five generations in Arrapha. 
 The cuneiform tablets listed in NPN are for contracts 
on land transactions, marriage, loans,  slavery, etc.  
In NPN, the  kinships and cuneiform tablets 
(contracts,  documents, texts)  involved are listed for each person. 
We reconstruct family trees from the added NPN to formulate the least squares problem with the constraints,   the father of a person is at least 22.5  years older than the person, contractors were living at the time of the contract, etc.   Our results agree with the Assyriological results of M. P. Maidman 
on the seniority among siblings of a powerful family. Our method 
could be applied to the other clay tablet archives once we have the name index in the format of NPN.
\end{abstract}

{\bf keyword}
 Nuzi Personal Names, family tree,  synchronism,  least squares, published year,  birth year,   


\section{Introduction}\label{introduction}
Radiocarbon dating \cite{laa} is having a profound impact on archaeology. It has many applications, including recent dating based on the Bayesian model for dynastic Egypt 
\cite{rd}. 
   Natural sciences, such as physics, chemistry, biology, etc.,  are effectively applied to archaeology \cite{a}. 
 Here, we study  mathematically and computationally Nuzi Personal Names (NPN) \cite {gpm}, and  the  Assyriological results of Maidman \cite{mai,  mai3} to  
   date  the Nuzi 
 cuneiform tablets and determine the birth years of persons appearing in the tablets.     Nuzi is a town from the 15th - 14th century B.C.E.
 (a period of some five generations (1475-1350 B.C.E.) \cite{mai3})  
 in Arrapha, in ancient Mesopotamia.  
 The year of publication is not written on any Nuzi cuneiform tablets. 
 Nuzi Personal Names (NPN), published in 1943 \cite{gpm},   is  an
 index of   personal names  in which  the kinships and 
the   cuneiform tablet  (contract, document, text) identifiers involved 
are listed for each person.
We   add the kinship informations in \cite{mai,  mai3} to 
electronic NPN   \cite{u2}, \cite{iium, umi, u, umit},  to reflect the studies after 1943 and derive an electronic file  RNPN2  \cite{u2} (Section \ref{adding}). 
We estimate published years of cuneiform tablets and the birth years of persons from RNPN2. 

The electronic files,  introduced in later sections, 
will be  given  in \cite{u2}. 

NPNelectronic\verb|_|data.txt,  

NPNelectronic\verb|_|correct\verb|_|data.txt,

 RNPN2\verb|_|data.txt, 
 
 RNPN2familytree.csv,  

NuzisamenamepersonIndex.csv

 We  reconstruct family trees from RNPN2 by using sequential algorithms  (Section \ref {criteria}), \cite{umit}, 
solve 
 name ambiguity problems \cite{hzg} 
 (Section \ref{RNPN2tree}, Section \ref{dispute})
and define the  Te\b{h}ip-tilla family  (a powerful family of 
the town of Nuzi) network of the 10253 persons 
  and 1665 documents related to the  Te\b{h}ip-tilla family (Section \ref{RNPN2tree}, Section \ref{inequality}).

We consider the following assumptions, (I), (II), and (III) (Section \ref{least}) for the persons and documents in the network for the generation length $g$.

 (I)
A father (mother) is 
at least $g$  years older than his (her)  children, 

(II)
A person appearing in a document is more than $\frac{2}{3}g$ years old.   
 Contractors, witnesses, and scribes were living at the time of the contract.   

 (III) The document  JEN 525 \cite{mai3} was published  
 in -1355 (1355 B.C.E.).   
\vspace{0.3cm} 
To decrease the number of parameters, we assume the age difference between mother and children is equal to that of father and children.
  JEN 525 contains 50 personal names, with the largest listed names
among the documents in the  Te\b{h}ip-tilla family network introduced in Section \ref{least}. We assume that JEN 525 was published 5 years after the destruction of Tursa by  Assyria in -1350, considering \cite{mai4}. 

We study  the  configurations   of birth years and death years of 10253 persons and published years of 1665  documents,  
which satisfy  the   
assumptions  (I), (II), and (III)) (Section \ref{least}.

Here we study:

1. The computationally obtained seniority among siblings for the persons in the Te\b{h}ip-tilla family as well as \b{H}i\v{s}meya family in Section 
\ref{hismeyasec} agrees 
with  Maidman's \cite{mai} Assyriological results (Section\ref{siblings},  Section\ref{hismeyasec} ). We show a
computational answer to the two problems (Section \ref{problem}) on seniority among siblings suggested by Maidman \cite{mai} and seniority among siblings.

2. The description (p. 253 \cite{mai3} "Overlap of witness among many of these documents suggests that these texts are written in close chronological  (and, of course, geographical: the town of Tursa is specifically named) proximity with each other. Substantial difference in the witness list suggest that proximity was not immediate. "  strongly suggests the computational study of the chronological order of documents. We study  the problem  in "The Decline and the Fall of a Nuzi Family" (Chapter Four of \cite{mai3})
in Section 
\ref{hismeyasec}.

 3. 
 A large  portion of the 
documents are on the land transaction, for example, 
 given
 later in the tablet with the 
document identifier   JEN 208 \cite{ch} (Section \ref{JEN208}) and Section\ref{hismeya}.  
Assuming  that the properties of the powerful family, as 
 Te\b{h}iptilla 
family,  are proportional to the accumulated number of documents for the 
family,  
the logistic growth \cite{umit} seems to show   the dynamics of 
 the concentration of properties  in  landowning by a few powerful families as the Te\b{h}iptilla 
family \cite{mai, mai2}.

In  p.49 \cite{po},
  ``The archives of the Mura\v{s}\^{u} firm at Nippur, consisting of
730 baked tablets and dating from 455 to 403 B.C.E., were discovered
in 1893 ".
The distribution of published 
year of 570 texts  in the  
  is given in the chart   (histogram)  
(Fig.3 Chapter V \cite{st}).  
  The logistic growth could be one of the reasonable models    
 to study the chart. 



Quantitative social network analysis is useful to study  the Mura\v{s}\^{u} archive \cite{wlk}.
We have another computational dating, for example, to the medieval English charters for the 11th - 14th-century \cite{tfg},    which 
utilizes the variation over time of word and phrase usage and distance between documents.

\section{Nuzi Personal Names}\label{NPN}
\subsection{Town of Nuzi}\label{town}
As in  the back cover of the book \cite{mai3}, 
``Ancient Nuzi, buried beneath modern Yorghan Tepe in northern Iraq, is a Late Bronze Age town belonging to the kingdom of Arrapha that has yielded between 6,500 and 7,000 legal, economic, and administrative tablets, all belonging to a period of some five generations (1475-1350 B.C.E.) and almost all from known archaeological contexts", 
see also \cite{starr,  mo}.
It was destroyed by  Assyria \cite{mai3}.

As in p. ix  NPN  \cite{gpm})``the name list, with its comprehensive data on genealogies and  professions, will permit
 assignment of the documentary 
sources to successive generations concerned. With the attainment of this basic 
chronological perspective the progress 
of legal, economic, and social change will become traceable."  
As in p.  1 and p.2  
of NPN \cite{gpm}``the tablets discovered at Nuzi belong to private  archives and official archives found in 
the houses of rich families  or to official archives kept in the palace. There is a wealth of
 texts pertaining to land transactions (buying, renting, exchanging), there are family contracts 
in the form of marriage documents and wills, there are transcripts of litigations and of 
declarations in court, there are loan tablets, slavery contracts, lists and inventions of 
tablets, and many other varieties. All these texts enable us to reconstruct the social 
and economic life of Nuzi 
in the middle of the 2nd millennium B. C. E. " 

As in  p. 127-128 \cite{mai} ``between twenty  
and twenty-five percent  deal  with the business transactions of a single family, 
that of Te\b{h}ip-tilla son of Pu\b{h}i-\v{s}enni."  
``Furthermore this group of texts represents  the  largest cuneiform family 
archive thus  far recovered from ancient Near East."

\subsection{ An example  of the Nuzi documents  for Nuzi Personal Names}\label{JEN208}
We have  4004 personal names in 1832 tablets with the names of books or periodicals in NPN
(i.e., AASOR, HSS, JEN, SMN, etc.). As given on line 25 from the bottom on p.8 of NPN,  the Arabic number 
following an abbreviation of a book or periodical title with or 
without a Roman numeral designating the volume is regularly that 
of a tablet in the publication, as in HSS V 30:5 or JEN 100:5. For example 
JEN is the abbreviation of the book Joint Expedition with the Iraq Museum at Nuzi.
As given on line 22 from the bottom on p.8 of NPN, the Arabic number after a colon represents 
the line number. 
  We should input (type) NPN carefully and faithfully to make NPNelectronic\verb|_|data. txt \cite{u2}.  For example, space, commas, and semicolons have important meanings in NPN. From the Assyriological viewpoint, remarks and corrections on NPN are shown in \cite{la}. From our computational viewpoint, 
 NPN has typographical errors, and there are also some mutually contradictory descriptions. We found out most of these errors and descriptions 
by chance in the process of 
programming. Although sometimes it is complicated to choose the correct description,  we choose one description to carry out our computation. 
We make NPNelectronic\verb|_|correct\verb|_|data.txt. The details are shown in \cite{u2}. We call NPNelectronic\verb|_|correct\verb|_|data.txt  electronic NPN (eNPN).  
We consider a  document with the document identified by JEN 208 to see how we 
use NPN for our computational studies   \cite{mak, iium, umi, u,umit}.
The document  JEN 208,  \cite{ch} \cite{gpm}, is 
the contract between  Ilu\b{i}a son of \b{H}amattar  and Te\b{h}ip-tilla  son of
 Pu\b{h}i\v{s}enni, with the names of witnesses and scribes.  
\vspace{0.2cm}

\noindent 
 {\bf JEN 208} \cite{ch}
(English translation \cite{mai5})
"Tablet of adoption of Iluya son of \b{H}amattar.  He adopted Te\b{h}ip-tilla son of Pu\b{h}i-\v{s}enni.
	Iluya gave to Te\b{h}ip-tilla as his inheritance share a 2.3 homer field-by the large standard-to 
the north of the dimtu (= district) of Inb=ilishu (and) to the east of the dimtu of Eniya.
	And Te\b{h}ip-tilla gave to Iluya as his gift 10 homers of barley.
	Should the field have claimants, Iluya shall clear (the field) and give it to Te\b{h}ip-tilla.
	Iluya shall pay the ilku (a kind of tax).
	(Then a mostly broken penalty clause. Only a part of the word ``abrogate" survives.)
    (The witness list is entirely destroyed.)
	Seal impression of Mar-i\v{s}htar son of Ataya; 	seal impression of It\b{h}ip-\v{s}harri son of Te\b{h}uya; seal impression of Kilip-\v{s}\b{h}ari son of Na\v{s}hwi; seal impression of [... son of ...]-arra, witness(?)."
\vspace{0.1cm}

 In NPN, personal names are listed alphabetically with the information 
in the original cuneiform documents (p. 6 of \cite{gpm}). The following example is
 on parts of NPN that are associated with two names, Ilu\b{i}a and \b{H}amattar, who appear in JEN 208. 
\vspace{0.2cm}

\noindent
 {\bf Example \ref{JEN208}.1.} 
On  \verb+ILUI_{n}A 1)+ in our electronic version \cite{u2}.
\vspace{-0.2cm}

{\tiny
\begin{verbatim}
H_{u}AMATTAR
 H_{u}a-ma-at-ta-ar, var. (2) H_{u}a-ma-at-ti-ir
  1) f. of I-lu-ia, JEN 208:2; (2) JENu 414; gf. of Ta-a-a, JEN 369:4
ILUI_{n}A
 Ilu-ia, var. (2) I-lu-ia
  1) s. of H_{u}a-ma-at-ta-ar, (2) JEN 208:1, 8, 11, 13, 14; 369:3, 10; H_{u}a-ma-at-ti-ir, (2) 
     JENu 414
  2) scribe, s. of ^{d}Sin-na-ap-s^{v}i-ir, JEN 226:42, 45; 438:21, 24
  3) s. of U^{'}-zu-ur-me, JEN 13:37
  4) f. of S^{v}a-ar-til-la, JEN 640:13; 662:95; HSS IX 7:31 (read so against Ili-iddina of copy); 
     35:38; RA XXIII 33:33; 50:43; 67:23
  5) f. of Ta-a-a, JEN 369:3, 10
  6) f. of Da-an-ni-mu-s^{v}a, JEN 345:5
  7) scribe, JENu 625; AASOR XVI 56:41
TAI_{n}A
 Ta-a-a, var. (2) Da-a-a, (3) Ta-a-ia, <<(4) Ta-ia>>
  30) s. of Ilu-ia, gs. of H_{u}a-ma-at-ta-ar, JEN 369:10, 14, 17, 18, 
    25, 26, 35, 36, 39, 42, 48
\end{verbatim}
}
For \b{H}amattar and  Ilu\b{i}a, in NPN, 
 we write the special character \b{H} as 
\verb+H_{u}+ and  \b{i} as \verb+i_{n}+ in electronic NPN (eNPN). 
For TA\b{I}A, with a special accent under the letter I,  in NPN,  
we write {\small TAI\verb+_{n}+A},  in the electronic NPN.
We have one person with the name \b{H}AMATTAR, as  {\small \verb+H_{u}+AMATTAR 1)},  seven  persons with the name ILU\b{I}A, as 
{\small \verb+ILUI_{n}A 1),..., 7)+}, and show them in the above. 
As we see by the argument in Section \ref{criteria}, the person {\small \verb+ILUI_{n}A 1)+} and the person {\small \verb+ILUI_{n}A 5)+}
represent the same person.
 We have 124 persons with the name  
 TAI\verb+_{n}+A as 
{\small \verb+TAI_{n}A  1),...,124)+}  in  NPN.   We show just  {\small \verb+TAI_{n}A 30)+}
 in the above.

As for \verb+H_{u}AMATTAR+   in the above Example \ref{JEN208}.1, 
in the 1st line, a personal name is given in capital letters as  
\verb+H_{u}AMATTAR+.  
In the second line, the variations of the name are shown.   
After the 3rd line, the kinship of the person with the name and  his (her)  
documents 
involved are listed as  
 JEN 208:2, which   shows that the person appears in line 2 of the cuneiform
 tablet JEN 208 (Arabic number after a colon represents 
the line number, as given on line 22 from the bottom on p.8 of NPN). In some cases, just the document identifier is shown 
without a line number, 
as JENu 414  (small letter u is for unpublished).

We have 12 persons in   JEN 208 listed with the line numbers in NPN. 
At  the contract JEN 208, we assume the following 7 persons out of 
the 12 persons who are listed in the form ``A son of B" and the persons without the  kinship information
were living.

{\scriptsize 
\begin{verbatim}
 208: 1 Ilu-ia, s. of H_{u}a-ma-at-ta-ar
 208: 3 Te-h_{u}i-ip-til-la, s. of Pu-h_{u}i-s^{v}e-en-ni
 208: 6 Im-bi-li-s^{v}u   
 208: 7 E-ni-ia  
 208:15 Ma^{n}r-^{d}is^{v}tar(U), s. of A-ta-a-a
 208:16 It-h_{u}i-ip-s^{v}arri, s. of Te-h_{u}u-ia
 208:17 Ki-li-ip-s^{v}e-ri, s. of Na-as^{v}-w<i>
\end{verbatim}
}
By our computation shown in Section \ref{introduction}, we estimate JEN 208 
was published in -1378 (1378 B.C.E.). 

\section{Criteria  to  identify persons in Nuzi Personal Names}\label{criteria}
We define   a standard representation  for the mutually different representations  of each 
personal name. 
 In many cases, a person is represented  by  
A  son of B, as 
Te\b{h}ip-tilla son of
 Pu\b{h}i\v{s}enni.  About 95\% of the names in NPN are male.
We should solve the following problems to identify persons. 
\begin{itemize}
\item
Problem 1. Many persons share the same name, for example, as we see 124 {\small \verb+TAI_{n}A 1),  2),..., 124)+}. 
\item
Problem 2. There are variations of a name in NPN. For example 
in  Example \ref{JEN208}.1, for  {\small \verb|H_{u}AMATTAR|,  
\verb|ILUI_{n}A|}  and {\small \verb|TAI_{n}+A|}.
\item
Problem 3. NPN  is  redundant from the viewpoint of 
family trees,   as 
we see in  
Example \ref{JEN208}.1. both  {\small \verb|H_{u}AMATTAR 1)|}, f. of   {\small \verb|ILUI_{n}A 5)|}  and  {\small  \verb|ILUI_{n}A 1)|}, s. {\small \verb|H_{u}AMATTAR 1)|}.
\end{itemize}
 In NPN, each person is listed by person identifier (person ID) given by a personal name with a number, 
as {\small \verb|H_{u}AMATTAR 1)|}, {\small \verb|ILUI_{n}A 1)|}, {\small \verb|ILUI_{n}A 5)|} in  Example \ref{JEN208}.1. Let us call the number,  the family number. 
We represent each person ID  in NPN  by the name number, in alphabetic order in NPN,  with the family number. For example 
we represent the  person ID   {\small \verb|ILUI_{n}A 1)|} in JEN 208 (Section \ref{JEN208}),  who has   
 the name number  1260  and the family number 1),  by $(1260,1)$.
  The  person ID  {\small \verb|ILUI_{n}A 5)|} is represented by  $(1260,5)$.

A family tree is a diagram showing the kinship between people in a family. 
We call the diagram showing the   kinships of   a person with a  person ID the  basic family tree of 
the person with the person ID. 
The  basic family trees of the four person IDs, 
{\small \verb|H_{u}AMATTAR 1)|,  \verb|ILUI_{n}A 1)|,   \verb|ILUI_{n}A 5)|} and  {\small \verb|TAI_{n}A 30)|} 
 (Example \ref{JEN208}.1), are four family trees represented by the kinships as, 
{\tiny 
\begin{verbatim}
 H_{u}AMATTAR
   1) f. of I-lu-ia,  gf. of Ta-a-a, 
 ILUIA
   1) s. of H_{u}a-ma-at-ta-ar, 
 ILUIA  
   5) f. of Ta-a-a, 
 TAI_{n}A
   30) s. of Ilu-ia, gs. of H_{u}a-ma-at-ta-ar,
\end{verbatim}
}

We say that each of the two family trees  $F_1$ and   $F_2$ is  
{\bf consistent}  with each other, 
 if and only if each of $F_1$  and $F_2$ shares at least   two names with each other    and does not have incompatible 
 kinships with each other for the shared names.   So, for the mutually 
consistent  $F_1$  and $F_2$, if $F_1$ 
 shares the names A and B with $F_2$, the kinship relation between A and  B  
 in $F_1$ should be compatible with the kinship relation between A and  B  
 in $F_2$.

 In Example \ref{JEN208}.1 the basic family tree of 
{\small \verb|ILUI_{n}A 1)|} and the basic family tree of {\small \verb|ILUI_{n}A 5)|} are not consistent with each other  when we consider just the basic family tree of {\small \verb|ILUI_{n}A 1)|}
 and the basic family tree of {\small \verb|ILUI_{n}A 5)|}.   
The following recursive sequential unifying algorithms   
 unify these four basic family trees into one family tree of the above four persons \cite{iium, u, umi}. The person  {\small \verb|ILUI_{n}A 1)|} and  the person {\small \verb|ILUI_{n}A 5)|} are considered  to be the same person by criteria 
   {\bf Criterion 0} shown later in this section.  
 \begin{itemize}
\item
{\bf Recursive sequential unifying} For the   stage $0\leq i$, we  put $T_i \equiv \{f_1, f_2,...,f_{n(i)}\}$,  as the initial set of family trees $S$.  
We apply the sequential unifying to $S$ and get $T_{i+1}$. 
 We apply the sequential  
 unifying,  by taking  $S=T_{i+1}$ as the initial set of family trees $S$ and   we get  $T_{i+2}$.    
 We continue the sequential unifying   until we get 
the first $i$ for   $T_{i-1}=T_{i}$,  put  $i=k$ and $T_k=T$.
\item
{\bf Sequential  Unifying}  
Let a set of family trees 
 $S_j\equiv\{f_j, f_{j+1}, ..., f_{n(j)}\}$ be obtained at the step $0\leq	j$ for $S_0=S$. 
Take the family tree $f_j$ and compare  
the family trees $f_{j+1},f_{j+2},...,f_{n(j)}$ sequentially 
with $f_j$.  
If  $k(j)$ family trees $f_{k(j)_1}, f_{k(j)_2},..., 
f_{k(j)_{k(j)}}\in F_j$
are consistent 
with  $f_j$
we unify  the  family trees $f_{k(j)_1},f_{k(j)_2},...,
f_{k(j)_{k(j)}}$ with $f_j$ 
 and get   the new unified  family tree $f_j $ 
from  the 
$k(j)+1$ family trees,   $f_j,f_{k(j)_1},f_{k(j)_2},...,
f_{k(j)_{k(j)}}$. 
 Put 
$T_j\equiv \{f_{k(j)_1}, f_{k(j)_2},...,f_{k(j)_{k(j)}}\}\in S_j$. 
We renumber the $ S_j \setminus T_j$ in the increasing order and get 
the set of the family trees 
as  $S_{j+1}\equiv \{f_1, f_2,...,f_{n(j+1)}\}$ where $n(j+1)=n(j)-k(j)$. 
We continue until $S_{m}\equiv \{f_1, f_2,...,f_{m}\}$,  and get the set of family trees 
$S \equiv S_m$.  
\end{itemize}

After the application of  the recursive sequential unifying, we can identify the persons by using the following criteria
\begin{itemize}
\item 
{\bf Criterion 0}.  If two persons X  and Y have the same name and if the basic family tree  of X and that  of  Y are consistent with  a
 the family tree in a set  of unified  family trees generated 
 at   $i$ (for $T_i$ in the above recursive sequential unifying ) and  
 $j$ 
(for $S_j$ in  the above  sequential unifying )   
X and Y  are regarded as the same person.  
\item
{\bf Criterion 1}.  
 If two persons, X and Y, have the same name and 
each of their names is given in the same line of the same document,  
then X and Y are regarded as the same person. 
\end{itemize}

\section{ Adding kinships and  identifying persons} \label{RNPN2tree}

 Adding  kinships to eNPN, we get the file RNPN2  \cite{u2} (
Section \ref{steps}). We reconstruct family trees identifying persons from RNPN2 and obtained the set of family trees, the file  $F_{RNPN2}$ \cite{u2}
 (Section \ref{steps}), which gives the linear inequality constraints Eq. \ref{ls2} by the assumptions, 
(Section \ref{inequality}, Section \ref{least}).

\subsection{Adding kinships to Nuzi Personal Names}\label{adding}

To study the Nuzi chronology      
computationally, we add the kinship information and the   
information on the identification of persons (Maidman  \cite{mai,  mai3}),
kinships obtained from the Te\b{h}ip-tilla family tree (Maidman\cite{mai,mai3}). (See the detail in \ref{subsectehip}.), 
kinships obtained from the  Kizzuk family tree (Dosch and Deller \cite{dd}, Maidman \cite{mai3}). (See the detail in \ref{subseckizzuk}.),  
the kinships given at the bottom of 
p 151 of the book \cite{mai3} (see the detail in Section \ref{dispute}, 
Identification of  persons from the information in
``A legal dispute over land: Two Generations of Legal Paper Works", from Chapter Three \cite{mai3}. (See the detail in 
Section \ref{dispute}).
\begin{figure} [ht]
\includegraphics[scale=0.25]{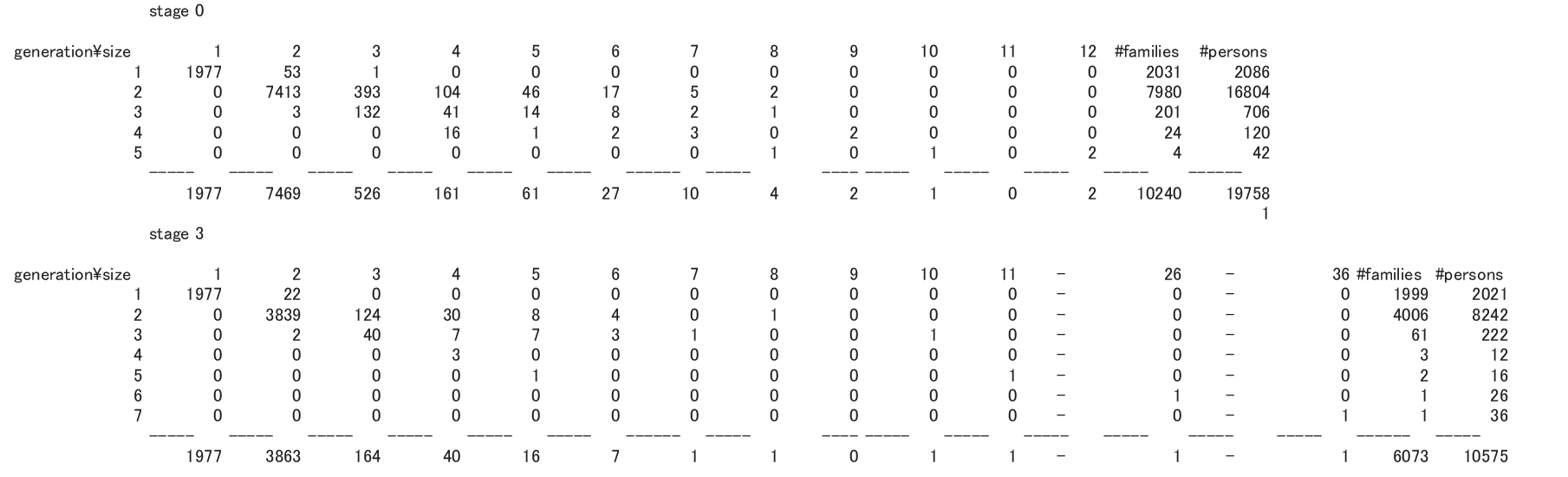}
\caption{Number of initial family trees classified by  size and height
and number of reconstructed family trees classified by  size and height}
\label{reconstructed}
\end{figure}
\begin{figure} [ht]
\includegraphics[scale=0.30]{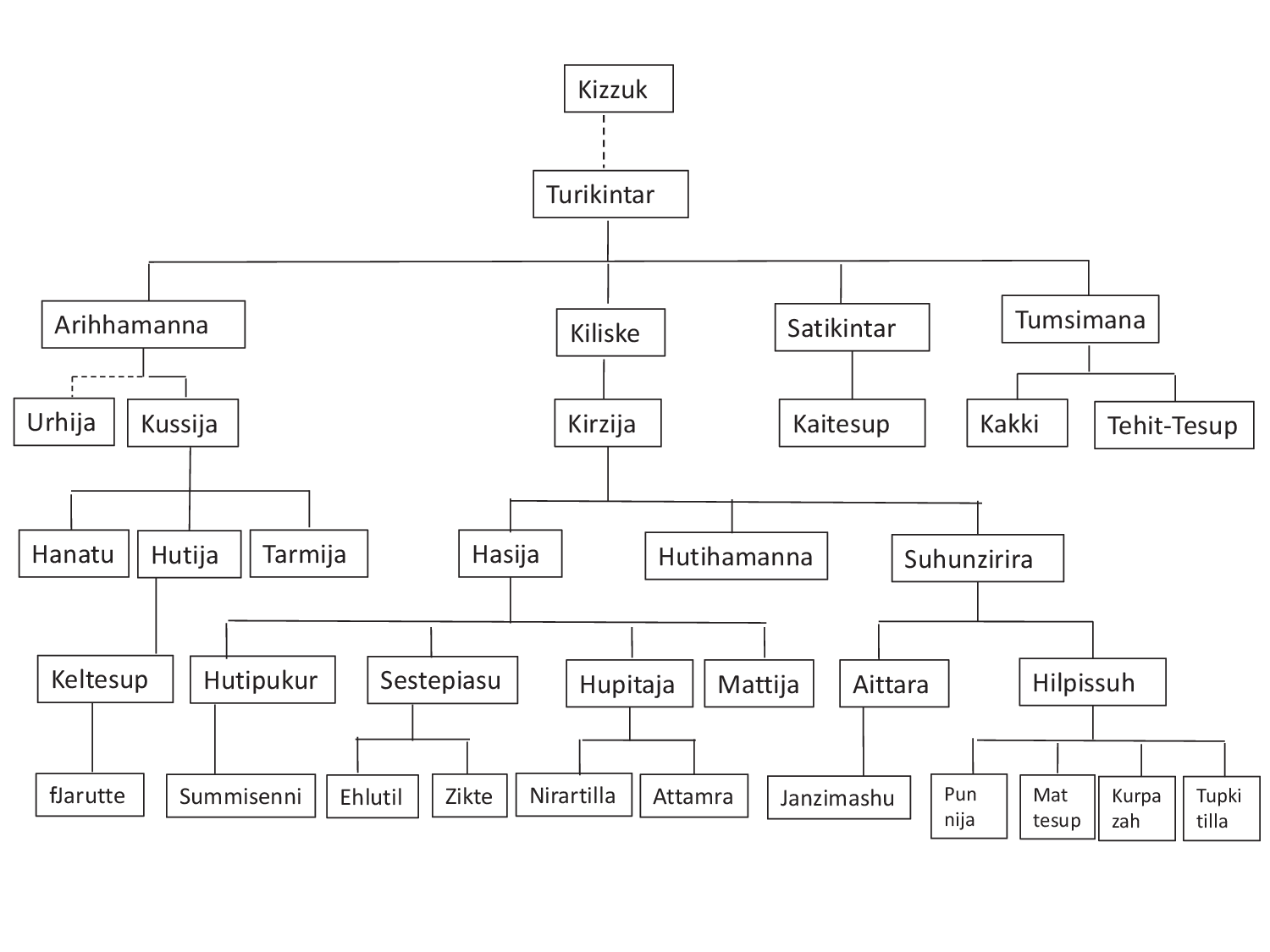}
\caption{ Kizzuk family tree   (Dosch and Deller \cite{dd} Maidman \cite{ mai3}).
 }\label{kizzuk}
\end{figure}

\subsection{Computational steps of adding kinships and identifying persons}\label {steps}
(1) We add  
the two kinships of the Te\b{h}ip-tilla family tree (Example \ref{subsectehip})
and 
the three kinships of the  Kizzuk family tree (Example \ref{subseckizzuk}.4) 
to eNPN to  make NPN2.  We do not add any person ID to eNPN. Hence, person IDs in eNPN and person IDs in NPN2 have one-to-one correspondence.

(2) We represent each person ID in NPN2  by the name number, in alphabetic order,  with the family number. We represent a person by a set of person IDs in NPN2. 
In Example \ref{JEN208}.1, we represent  the person with person ID  ILUIA  1) in JEN 208 (Section \ref{JEN208}),  who has   
 the name number  1260  and  the family number 1),  by 
 $\{(1260,1)\}$. 
 
  We represent the person with the  $(1260,1)$ by  the set $\{(1260, 1)\}$.  
Let $V_{NPN2}$ be the set of all persons in NPN2. We have 
$\{(1260, 1)\}\in V_{NPN2}$.  Each person has his/her set of kinships ( basic family tree) and his/her set of listed documents, as we see in Example \ref{JEN208}.1.

(3) We apply  the recursive sequential unifying (Section \ref{criteria})  to the set 
of persons  $S_{NPN2}$  of NPN2  and identify the persons   by using {\bf Criterion 0} to  get the set of persons $V_{NPN2.0}$. 
 (See Section \ref{enna}.)

(4) To decrease the cases in which 
   plural persons represent the same person. We apply {\bf Criterion 1} in Section \ref{criteria} to check all lines of all the documents listed in NPN2  sequentially and get the set of persons $V_{NPN2.1}$.  
(See Section \ref{enna}.)

(5) Consider $V_{NPN2.0}\cup V_{NPN2.1}$. 
The set of person IDs of each person  in $V_{NPN2.0}$ 
is disjoint with the set of person IDs of the other person in $V_{NPN2.0}$. For each person in $V_{NPN2.0}$, we sequentially identify with the persons in   $ V_{NPN2.1}$   that two persons are the same person if and only if there is at least one shared person ID.   Then we have $V_{NPN2.2}$.
(See Section \ref{enna}.)

(6)  We add the information in Section \ref{adding}
We put the kinships and the identifications  given in Section \ref{adding} to $V_{NPN2.2}$, 

\begin{tabular}{ll}
{\small  \verb|AITTARA|}     &(42,1),  (42,4) \\
{\small  \verb|ARIH_{u}-H_{u}AMANNA|}    &(314,    13),  (314,15), (314,20),    (314,25)\\
{\small  \verb|AS^{v}TAR-TILLA|}   &(560,     1),   (560,6) \\
{\small  \verb|H_{u}UTII_{n}A|}    &(1158,    23), (1158,    65), (1158,    92)\\
{\small  \verb|TURI-KINTAR|}    &(3467,     1), (3467,     2),    (3467,     3),  (3467,     4) \\
{\small  \verb|^{f}WINNIRKE|}   &(3779,     1), (3779,     2), (3779,     3)
\end{tabular}

\vspace{0.1cm}
\noindent
and others. Then we have $V_{RNPN2}$, \cite{u2}.

(7)  We put all the kinship information and the document information of each person in $V_{RNPN2}$.  Then we  get RNPN2  \cite{u2}). 
 For example, in RNPN2,  the persons 
$\{(1260,1)\}$ and , $\{(1260,5)\}$ are the same person (Section \ref{criteria}),   given by  $\{(1260,1),(1260,5)\}
\in V_{RNPN2}$ and written with the kinship information and document information  as,     
{\tiny
\begin{verbatim}
ILUI_{n}A
 Ilu-ia, var. (2) I-lu-ia
  1) s. of H_{u}a-ma-at-ta-ar, (2) JEN 208:1, 8, 11, 13, 14; 369:3, 10; H_{u}a-ma-at-ti-ir,
   (2) JENu 414, f. of Ta-a-a, JEN 369:3, 10
\end{verbatim}
}

The distribution of the size and the generation (height)  of the set of basic family trees    $S_{RNPN2}$ of RNPN2 is shown 
in  Fig \ref{reconstructed}.

 (8) We apply the recursive sequential unifying algorithms (Section \ref{criteria})  to the set of basic family trees  $S_{RNPN2}$ of RNPN2 and  get the set of unified family trees $F_{RNPN2}$,
  \cite{u2}.  Statistical distribution on size and height of family trees  of $F_{RNPN2}$ 
is in Fig \ref{reconstructed}. 
 
 (9) From $F_{RNPN2}$, considering RNPN2, 
we obtain the linear  inequality  constraints (Section \ref{inequality}, Section \ref{least} ) for the least squares problem.

The table, number of reconstructed family  trees classified by size and height, in Fig \ref{reconstructed} 
is for the set $F_{RNPN2}$. 
We have $k=3$ for the recursive sequential unifying in Section \ref{criteria} to get  $F_{RNPN2}$ from $S_{RNPN2}$.

Remark 1.
We obtained 
the Te\b{h}ip-tilla family tree of 26 members in NPN using the Te\b{h}ip-tilla family tree 
\cite{mai, mai3} of 36 persons.

\subsection{Adding kinships of the Te\b{h}ip-tilla family}\label{subsectehip}

 Pu\b{h}i-\v{s}enni,  who is the  father of Te\b{h}ip-tilla, is a son of Tur-\v{s}enni 
 (\# 84 JEN 552 \cite{mai3}). 
$^f$Winnirke 1), 2)
of NPN are identified as the same person by our computation.
  $^f$Winnirke 3),  in JEN 552, of NPN is the same person as the person 
$^f$Winnirke 1), 2) (mother of Te\b{h}ip-tilla son of Pu\b{h}i-\v{s}enni),   (note 104 for JEN 552, in p.258 \cite{mai3}).

{\tiny 
\begin{verbatim}									
^{f}WINNIRKE
 ^{f}Wi-in-ni-ir-ge, var. (2) ^{f}Mi-ni-ir-ki, (3) ^{f}Wi-in-ir-ge, (4) ^{f}In-ni-ir-ki, 
 (5) ^{m}Wi-ni-ir-ki, (6) ^{m}Mi-in-ni-ir-ki
  1) m. of H_{u}a-i-is^{v}-te-s^{v}up, gm. of Um-bi-ia and Ar-ru-um-ti, JEN 324:7, 12, 25, 29
  2) m. of Te-h_{u}i-ip-til-la, gm. of En-na-ma-ti, S^{v}ur-ki-til-la, and 
    A-kip-[ta-s^{v}e-en-ni], JEN 324:7, 12, 25, 29
  3) JEN 82:1, 3, 8, 10, 27; (4) 164:5; (3) 504:2; (2) 552:10; (2) 560:2, 9, 17, 24, 30, 36, 
    42, 48, 54, 60, 66, 73, 79, 85; (2) 561:3, 7, 11, 14, 19, 23 (no det.); (5) 562:3, 7, 8, 
    10, 14, 17, 18, 22, 23, 25, 27, 28, 30, 31; 575:9; (6) 647:2
 \end{verbatim}
}

\begin{figure} [ht]
\includegraphics[scale=0.30]{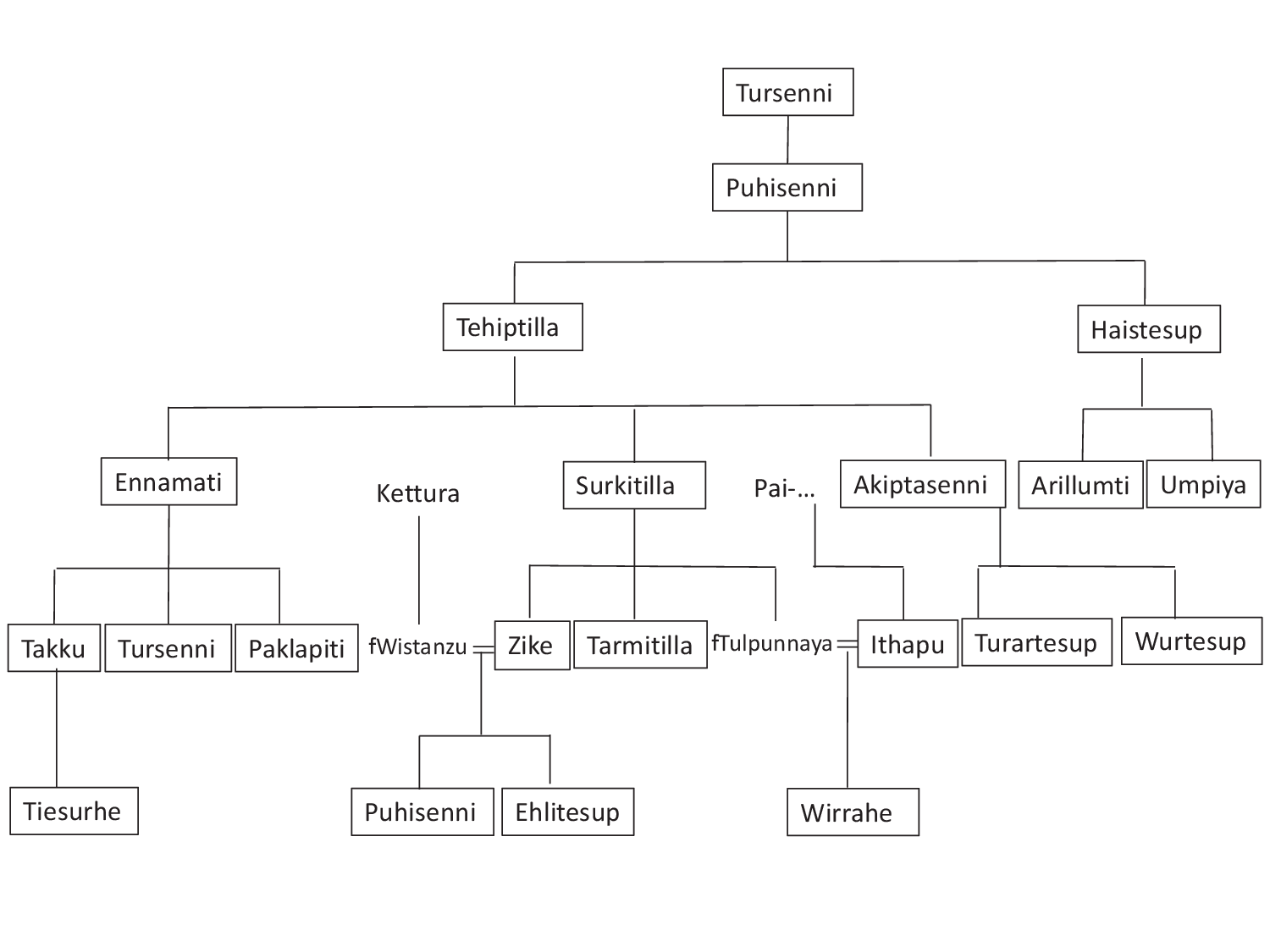}
\caption{Te\b{h}ip-tilla family tree for persons of NPN in Fig \ref {tehipfamily}, a part of  the family tree of Maidman \cite{mai, mai3}.
 }\label{tehip3}
\end{figure}

\vspace{-0.5cm}
\noindent
We  identify the 26 persons in NPN  mainly by  ``A son of B" from the  Te\b{h}ip-tilla family tree 
\cite{mai, mai3}. 
We  add the  following two  kinships  from the  Te\b{h}ip-tilla  family tree  \cite{mai, mai3}, and we have 
the Te\b{h}ip-tilla
  family tree Fig \ref{tehip3} of 26 members  by our computational algorithms (Section \ref{RNPN2tree}). 
\vspace{0.2cm}

\noindent
{\bf Example \ref{subsectehip}} Two additional kinships for Te\b{h}ip-tilla family tree of 26 members.

\vspace{-0.1cm}
{\tiny
\begin{verbatim}
S^{v}URKI-TILLA			
  12) f. of It-h_{u}a-a-p<u>			
WIRRAH_{u}H_{u}E			
  3) gs of S^{v]ur-ki-til-la			
\end{verbatim}
}

\subsection{Adding kinships of the Kizzuk family }\label{subseckizzuk}
We identify each of the 36 persons in the 
 Kizzuk family tree (\cite{dd, mai3}) with a person in NPN mainly in the form of A son of B 
(or B father of A).  
Assuming the Kizzuk family tree, we have the following Examples.

\vspace{0.2cm}

\noindent
{\bf Example \ref{subseckizzuk}.1.}  The following two persons are the same person since their basic family trees are consistent	with the Kizzuk family tree \cite {dd,mai3}
{\tiny
\begin{verbatim}
AITTARA
 A-i-it-ta-ra, var. (2) A-i-it-ta-a-ra, (3) A-it-ta-ra, (4) A-i-da-ra, (5) At-ta-ra
  1) s. of S^{v}u-h_{u}u-zi-ri-ru, (1)(4) JEN 604:17, 27; [S^{v}u-h_{u}u]-un-[z]i-ri-ra, 
    JENu 423; S^{v}u-uh_{u}-ni-zi-ru, (2) AASOR XVI 37:36
  4) f. of Ia-an-zi-ma-as^{v}-h_{u}u, (3) AASOR XVI 24:9; (3) 30:31; (3) 32:27; (3) 34:33; 
\end{verbatim}
}

\noindent
{\bf Example \ref{subseckizzuk}.2.} The following four persons are the same person since their basic family trees are consistent	with the Kizzuk family tree \cite {dd,mai3}.
{\tiny
\begin{verbatim}											ARIH_{u}-H_{u}AMANNA																	
  13) s. of Du-ri-ki-in-tar, JEN 312:22, 30; JENu 917; Tu-ri-ki-tar, (2) JEN 474:37, 49; 
    Du-ri-ki-tar, (2) JEN 554:37; s. of Tu-ri-[ki-in-tar], br. of S^{v}a-te-ki-in-tar, 
    JEN 232:25 
  15) f. of A-kap-s^{v}i-ia, (2) JEN 471:2; read Ku*-us^{v}*-s^{v}i-ia 
  20) f. of Ku*-us^{v}*-s^{v}i-ia, (2) JEN 471:2 (read so against A-kap-s^{v}i-ia of 
    copy---PMP) 						
  25) f. of Ur-h_{u}i-ia, JEN 342:2																\end{verbatim}
}

\noindent
{\bf Example \ref{subseckizzuk}.3.} {\small \verb|TURI-KINTAR 1), 2), 3), 4)|} are the 
same person since their basic family trees are consistent	with the Kizzuk family tree \cite {dd,mai3}.

\noindent
{\tiny
\begin{verbatim}
TURI-KINTAR	
  1) f. of A-ri-h_{u}a-ma-an-na, (4) JEN 232:24; 312:22; JENu 917; Ar-h_{u}a-ma-an-na, (2)
    JEN 474:37; (3) 554:37	
  2) f. of Ki-li-is^{v}-ge, JEN 486:36	
  3) f. of S^{v}a-ti-ki-in-tar, (4) JEN 90:10, [20]; HSS V 48:4; 49:33; S^{v}a-te-ki-in-tar,
    (4) JEN 232:24; S^{v}a-di-ki-in-tar, (3) JENu 323; S^{v}a-te-ki-tar, (2) HSS V 47:41	
  4) f. of Tu-us^{v}-ma-na, (2) JEN 79:18; Du-um-s^{v}i-ma-na, JEN 644:5	
\end{verbatim}
}																

\noindent
After the identification mainly by  ``A son of B", 
we  add the  three  kinships Section \ref{steps} (1) obtained from the  Kizzuk family tree  \cite{dd, mai3}, then we have Kizzuk family tree of 36 members Fig \ref {kizzuk} by our computational algorithms (Section \ref{RNPN2tree}).

\vspace{0.2cm}
\noindent
{\bf Example \ref{subseckizzuk}.4.}
 Three kinships in the Kizzuk family tree \cite{dd, mai3} to construct a Kizzuk family tree of 36 members Fig \ref{kizzuk}.
{\tiny 
\begin{verbatim}							
AITTARA				
  1) gs. of Ki-ir-zi-ia				
H_{u}ILPIS^{v}-S^{v}UH_{u}				
  1) gs. of Ki-ir-zi-ia				
KUS^{v}S^{v}II_{n}A				
  13) gf. of Ge-el-te-s^{v}up
\end{verbatim}
}

\noindent
We add {\small \verb|AITTARA 1) gs. of Ki-ir-zi-ia|}, in the above 
three kinships in NPN from the Kizzuk family tree
\cite{dd, mai3},  to the   {\small \verb|AITTARA 1)|}.	
{\tiny 
\begin{verbatim}
AITTARA
 A-i-it-ta-ra, var. (2) A-i-it-ta-a-ra, (3) A-it-ta-ra, (4) A-i-da-ra, (5) At-ta-ra
  1) s. of S^{v}u-h_{u}u-zi-ri-ru, (1)(4) JEN 604:17, 27; [S^{v}u-h_{u}u]-un-[z]i-ri-ra, 
    JENu 423; S^{v}u-uh_{u}-ni-zi-ru, (2) AASOR XVI 37:36
  2) s. of U^{'}-ge, JEN 468:39 (last sign ra accidentally omitted in copy)
  3) f. of Ge-li-ia, JEN 600:28
  4) f. of Ia-an-zi-ma-as^{v}-h_{u}u, (3) AASOR XVI 24:9; (3) 30:31; (3) 32:27; (3) 34:33;  
    I-in-si-ma-as^{v}-h_{u}u, (5) JEN 87:2
  5) scribe, RA XXIII 32:40, seal
  6) (3) AASOR XVI 22:18; 31:28; 45:19 (read so against I-it-ta-ra of translit.)
\end{verbatim}
}  
\noindent
In the same way, we add the other two kinships to NPN.

 We have 36 members in NPN in Fig \ref{reconstructed}
 out of the 37 members \cite{dd, mai3} of the Kizzuk family tree Fig \ref{kizzuk} by the argument in Section \ref{RNPN2tree}. 
  We can not find in NPN, \cite{gpm}, published in 1943,  the person   Attamra son of Hupitaja, in the Kizzuk family tree \cite {dd, mai3}. He is listed  in HSS XIX 9 \cite{cg} 
, published in 1977, (\cite{mai6}
). 
The Kizzuk family tree \cite {dd, mai3} gives a good example of how parts of a big family tree are given in NPN.

\subsection {For  ``A legal dispute over land" (Chapter Three \cite{mai3}) and  "the decline and fall of a Nuzi family" (Chapter Four \cite{mai3}}\label{dispute}
For Chapter Three   we put 	
the kinships given at the bottom of 
p 151 of the book \cite{mai3}, `` [\b{H}i\v{s}meya son of It\b{h}i\v{s}ta, of] 
{\small \v{S}arra-\v{s}ad\^{u}ni}(??) [son of] It\b{h}i\v{s}ta(?), 										
and of U\v{s}\v{s}en-naya daughter of [Enna-milki] (and) wife of It\b{h}i\v{s}ta".	
 (JEN 603  \cite{mai3} ).

Chapter Four  has seven documents.  All seven documents 
 are in NPN, and we can estimate the published years. 

\# 55 JEN 644 (\cite{mai3}) should be the start of the legal dispute (Chapter 3     \cite{mai3}).  
  NPN needs more information to connect kinships using our criteria and algorithms \cite{u, umit}. 
We add the following information. 
  Then we have the estimated years, which do not contradict the results \cite{mai3} except \#58 JEN 512 and \#59 JEN 135.

\begin{enumerate} 
\item 
The following two persons {\tiny \verb|H_{u}UTII_{n}A 23)|} in JEN 644 and

\noindent 
{\tiny \verb|H_{u}UTII_{n}A 65)|} in JEN 321 in NPN should be identified to the same person. 
(Kizzuk family tree \cite{mai3, dd}).  

{\tiny
\begin{verbatim}
H_{u}UTII_{n}A
  23) s. of Ku-us^{v}-s^{v}i-ia, JEN 118:23 (read so against Ku-us^{v}-s^{v}i-i of copy);
    119:6, 11, 16; 342:1, 4, 11, 13, 19, 25, 31, 36, 40, 41; (1)(2) 364:2, 21, 25, 27,
    31, 33; (2) JENu 623; HSS IX 8:1, 4, 7, 32; 12:1, 5, 8, 37; 17:19, 26; 18:44, 59;
    41:6; 43:4, 15, 18; TCL IX 40:6; br. of Tar-mi-ia, JEN 666:1, 11, 16, 17, 36, 37, 46;
    Gu-us^{v}-s^{v}i-ia, JEN 218:12, 21; 600:2, 9, 13, 15, 17; Ku-s^{v}i-ia, JEN 644:10,
    12, 15, 22; JENu 991
  65) f. of Ge-el-te-s^{v}up, JEN 29:3; 83:4; 85:3, 8, 10; 111:5; 116:7; 125:1; 143:8;
    186:6; 219:3; (2) 316:4; 321:1; 340:1; 469:2; 477:2; 541:2; 592:3; 602:3; 616:2; 
    646:4; 672:1, 4, 43, 47; JENu 620; 703; 729; (2) 730; gf. of ^{f}Ia-ru-ut-te, JEN 435:2
  92) (1)(2) JEN 59:3, 11, 13, 14, 21, 25; 117:4, 6; 154:24; 233:24; 254:24; 325:4, 7; 388:1,
    6; 495:6; 531:23; 534:5; 627:4; JENu 65; (2) 255; 288; 485; 533; 624; 1056; (1)(2) 1168
\end{verbatim}
}
in \# 56 JEN 388 and \# 57 JEN 325 should be the above {\small \verb|H_{u}UTII_{n}A| 23)} guessing from the name list JEN 644, JEN 388 and JEN 325 in Chapter three \cite{mai3}. 

\vspace{0.3cm}
\item 
{\tiny \verb|AS^{v}TAR-TILLA| 1)} and {\small \verb|AS^{v}TAR-TILLA| 6)}
 should be the same person guessing from the name list of JEN 644 and JEN 388 Chapter Three\cite{mai3}. 
{\tiny
\begin{verbatim}
AS^{v}TAR-TILLA (see also Artar-tilla)
 As^{v}-tar-til-la
  1) s. of Pu-i-ta-e, JEN 600:40; 644:26, 36; br. of Ut-ta-zi-na, JEN 59:33, 38
  2) shepherd, s. of Du-ra-ri, HSS IX 26:1, 17
  3) s. of [....]-u^{~}, HSS IX 152:rev. 10
  4) f. of Ur-h_{u}i-til-la, HSS IX 35:11
  5) mas_{.}s_{.}artu, HSS IX 37:10
  6) JEN 388:1, 7
\end{verbatim}
}
\end{enumerate}

 \subsection{ Identifying Enna-mati son of Te\b{h}ip-tilla} \label{enna}
  {\small 
 The 8 persons in NPN2, 
 
\noindent
 $\{(744,    16)\}, \{(744,     67)\}, 
\{(744,   44)\},  \{(744,     61)\},\\ \{(744,      71)\},
\{(744,       79)\},\\ \{(744,      80)\}, 
\{(744,       85)\}\in V_{NPN2}$  are identified to the same person 
 as follows.
 
 Let us consider Enna-mati son of Te\b{h}ip-tilla 
to understand the  computational steps   (2), (3), (4), (5)
 and (6) in Section \ref{steps}). 
 
(2) The personal  name {\small \verb|ENNA-MATI|} has  family numbers,   
1),..., 85).  
 The person with person ID {\small \verb|ENNA-MATI 44)|},  
 is represented by the set of the person ID $\{(744, 44)\} \in V_{NPN2}$.

(3) 
We  apply  the recursive sequential unifying to NPN2   to
identify by 
{\bf Criterion 0}.  We  find that the persons 
$\{(744,16)\}, \{(744,  44)\}, \{(744,    67)\}, \\
\{(744,      79)\},
\{(744,       80)\}\in V_{RNPN2}$ are identified to the same person and written as the set of the person IDs as \\ 
$\{(744,16), (744,  44), (744,    67), (744,      79),
(744,       80)\} \in V_{RNPN2.0}$.    

(4) 
We
sequentially apply 
{\bf Criterion 1}  to NPN2  for   and get
from the same line of the same document, JEN   655 line- 1,                               
$\{(744,       16),(744~      44)\}\in V_{NPN2.1}$, and 
  from   the same line of the same document JEN   515  line-7                                
$\{(744,         79),(744~      85)\}\in V_{NPN2.1}$.

(5)
We see $\{(744,  16),(744, 44),(744, 67),(744, 79), (744, 80)\}\in   V_{NPN2.0}$, 

\noindent
and
$\{(744, 16),(744, 44)\},\{(744, 79),(744 , 85)\} \in V_{NPN2.1}$.

We obtain 

\noindent
$\{(744,  16),(744, 44),(744, 67),(744, 79), (744, 80)
,(744 , 85)\}\in   V_{NPN2.2}$.

(6) 
From the Te\b{h}ip-tilla family tree (Maidman \cite {mai, mai3} )

\vspace{-10pt}
{\footnotesize
\begin{verbatim}
ENNA-MATI  
  61) f. of Pa-ak-la-bi-ti, JEN 525:20; 670:23; Ba-ak-la-bi-ti, JENu 354
  71) f. of Dur-s^{v}e-en-ni, JEN 335:1; JENu 164
\end{verbatim}
}
\noindent
, represented by  $(744, 61)$ and $(744, 71)$, 
are identified to \verb|ENNA-MATI| son of \verb|TEH_{u}IP-TILLA|	$(744, 44)$.
Thus 8 persons are identified to the person 

\noindent
$\{(744,   44), (744,  16), (744,       67),(744,       79),$
$(744,     80), (744,      85), (744,     61), (744,     71)\}$
$\in V_{RNPN2}$. 
}


\section{Linear inequality constraints obtained from reconstructed family trees}\label{inequality}
As we see in Fig. \ref{reconstructed}, we have 4096 family trees of plural members in  the set $F_{RNPN2}$, the total number of members of the 
4096 family trees are 8598. We have 1977 single-family trees.
 Hence we have 8598+1977=10575 persons. 
We are interested in the persons who lived at the time of each contract. 
Hence we do not count the persons just listed as f. of, m. of, gf. of. 
We assume  A  son of  B   in a document lived at the time of contract of the document.  
The assumption does not give information on the life or death of  B   at the time of  
the contract. For example we assume  {\small \verb|ILUI_{n}A 1)|} lived at the time of the contract 
JEN 208.  However, the assumption does not mean   the life or death of {\small \verb|H_{u}AMATTAR 1)|} 
 at the time of the contract JEN 208.  
 ``A s.  of B" means  ``B  f.  of A" and  gives the same inequality on age  to 
``B  f. of A".   
 ``A gs.  of B" means  ``B  gf.  of A"  gives the same inequality to ``B  gf. of A".    
 Hence from each of the 10575    basic family trees,   $S_{RNPN2}$,  
  we delete the basic family trees  with  
 ``f.  of A",  ``gf. of A", 
 ``ggf. of A",  ``m. of A",  ``gm. of  A",  ``ggm. of  A" to consider the inequalities for the above 
  (I) and (II).

Then we have 6489 persons. 6288 persons out of 6489 persons are in 
the Te\b{h}ip-tilla family network. Thus we have
persons 1,2,...,6141, and persons 6142,6143,...,6288 who are not identified persons, as the following person [....]-ia
\noindent
{\tiny 
\begin{verbatim}
ARIM-MATKA
  16) f. of ....-ia, (2) JEN 618:32
\end{verbatim}
}

\noindent
in NPN. 6289 - 10253 are added persons who appear as f. of , m. of ,
gf. of the 6288 persons.

We assume persons in one-generation family trees, mostly one-person family trees, were living at the time of publication of the listed documents.

In a document, we assume  A son of B, was present at the time of the publication  (A son of B assumption).  
Here we apply the ``A son of B" assumption to the Te\b{h}ip-tilla family tree of 26 persons.
 For the other family trees of two or more generations,    we assume  the persons who are  at  the 
top of the family tree,  are not present at the publication. 
We  approximate   the 
A son of B assumption by this assumption, since we see, from Fig \ref {reconstructed},  2021 persons are in the family trees of one generation,  8242 persons are in the family trees of two generations, and 26 persons are the Te\b{h}ip-tilla family tree.   The remaining 286 persons, out of 10575 persons,  are for the object of the approximation.  We assume the approximation  does not 
seriously affect our results on Te\b{h}ip-tilla family.

We need this approximation since we can not apply simple algorithms  to each of 6073 family trees (Fig 1).  For example we have  7  mutually contradicting descriptions
 in NPN \cite{u2},  as

{\tiny
\begin{verbatim}
ARIH_{u}-H_{u}A
  3) s. of En-na-ma-ti, (2) JEN 118:28; (2) 174:14; JENu 1040 
(see also Arih_{u}-h_{u}amanna, f. of idem)
ENNA-MATI
  2) s. of A-ri-h_{u}a, JEN 118:28; 174:14; A-ri-ih_{u}-h_{u}a, 
JENu 1040 (see also following number)
\end{verbatim}
}

\noindent
which prevent making a system of inequalities, for A son of B  assumption, without contradictions. 
We observe that most of the persons in NPN are in a one-person-family tree or 
in a two-person-family tree  \cite{u, umit}. 
 45\% of the  family trees share documents with  Te\b{h}ip-tilla family.
17 families with 
79 persons share 70\% of documents.

\section{   The  least squares with constraints  for the
Te\b{h}ip-tilla family  network}\label{least}


 Let us define the Te\b{h}ip-tilla family network (Section \ref{introduction})  rigorously.  A document of the Te\b{h}ip-tilla family is defined as a document in which at least one person of Te\b{h}ip-tilla family of the 26 members appears. 
Two documents, A and B, are connected if the list of persons appearing in  A and those appearing in  B  have at least one shared person. 
A document of Te\b{h}ip-tilla family is a document in which at least one person of Te\b{h}ip-tilla family of the 26 members appears.
A document $A$ ($A\equiv B_0$) is in the  Te\b{h}ip-tilla family network 
if and only if there is a sequence of documents  $B_i$  $i=0,1,...,n-1$,  where $B_i$ is connected with $B_{i+1}$ for  $i=0,1,...,n-1$  and $B_n$ is a document of   the
 Te\b{h}ip-tilla family. 
We have 10253 persons  and 1665 documents in
 the Te\b{h}ip-tilla family  network.
Each of the 10253 persons $i=1,...,10253$, with the birth year $b_i$ and death year $d_i$,  belongs to one family tree, and each person appears in one or more documents.   One or more persons appear in each document,  $P_k$, $k=1,...,1665$,  with the published year ( the year recorded as a cuneiform tablet).    
From RNPN2 (Section \ref{adding}) we obtain  39782 linear inequalities  Eq. \ref{ls2} for the assumptions (I) and  (II).  
\begin{equation}\label{ls2}
  \left\{
   \begin{array}{ccl}
b_i &\geq & b_{f_i} +g\\
b_i &\geq & b_{m_i} +g\\
d_i &\geq &   P_k \geq  b_i +\frac{2}{3}g \\ 
   \end{array} \right. 
   \end{equation}
and
   \begin{equation}\label{jen525}
   P_{k_0}=-1355 ~~\mbox{where}~k_0 ~\mbox{is the $k$ of JEN~525},
\end{equation}
where the person $f_i$ is the father  
of the person $i$ and  $m_i$ is the mother   
of the person $i$.

  We generate 
the  uniform random numbers to simulate  a set of 10,253 mutually independent     
 on the interval $[g,4g]$  for the lifespan   $U_i$   of the person $i$,   $i=1, ..., 10253$ of the Te\b{h}ip-tilla family network ( Section \ref{simulation}).

We wish to have the  configurations of the above birth years $b_i$s,   death years  $d_i$,  and published years $P_k$ in the Te\b{h}ip-tilla family network, 
which satisfy  the 39783 constraints, Eq. \ref{ls2} and Eq. \ref{jen525} (  
assumptions  (I), (II), and (III)).
Among the configurations
we take the one which is the closest to the $U_i$, $i=1,...,10253$, that is to say, which  gives the least squares, 
\begin{equation}\label{objective}
\sum_i((d_i-b_i)-U_i)^2,
\end{equation}
with  22171 
(= 10253 $\times$ 2 
(representing $b_i, d_i $ ) + 1665 
(representing $P_k$)) variables and 39783 constraints,   Eq. \ref{ls2},and  Eq. \ref{jen525} by using  NUOPT \cite{nuopt, sm, yyt}. 
Although the objective function Eq. \ref{objective} does not include $P_k$, 
 the variables $b_i,~d_i,~P_k$ interact with each other in the optimization process to satisfy  39783 constraints.

\subsection{ On simulated life spans}\label{simulation}

To see the stability of our numerical optimization,
we generate 10253 uniform random numbers 10 times as a simulation of life spans
$U_1(r),..., U_{10253}(r)$, $r=1, ..., 10$,   for $g=22.5$.  
We solve the least squares problem  10 times for each $r$  to obtain $b^0_i(r)$, $d^0_i(r)$ and $P^0_k(r)$.
 The birth year $b^0_i(r)$ of Te\b{h}ip-tilla family members is shown in Figure \ref{tehipfamily} for $r=1,...,r$ with their mean.

  As we can see from Eq. \ref{ls2}, there is no inequality to bound $d_i$ from above, while there are inequalities to bound $d_i$ below.
Persons with relatively many constraints,  as in the Te\b{h}ip-tilla family and 
 in the Kizzuk family,  are sometimes in the case $d^0_i-b^0_i>U_i$  depending on random numbers.   
 Otherwise, our computation by NUOPT shows
$d^0_i-b^0_i=U_i$. 
The case $d^0_i-b^0_i<U_i$ never occurs.

 We wish to have a configuration which satisfies the assumption (I), (II), and (III)
From a computational viewpoint, 
  the uniformly distributed life span on the interval $[g,4g]$ is the most reasonable among the models on intervals  $[g, mg]$, $m=2,3,4$.   
The  arithmetic mean   of 
 the least squares of 100 computations (optimizations) for the model on  $[g,4g]$ is the smallest among those on $[g,2g]$,  $[g,3g]$, and $[g,4g]$, with the respective numerical values,  
 52227.5, 25811.8, and 14880.6.
 We assume $\frac{2}{3}g$ in (II).  It seems more natural than assuming $g$ in (II) in the sense of the least squares.    
If we take   $g$ instead 
of $\frac{2}{3}g$  in assumption (II)  for the above three cases $[g,2g]$,  $[g,3g]$, and $[g,4g]$, 
the arithmetic means of the least squares are   114187.3, 59205.4, and  35775.5, respectively and larger than the above numerical values for $\frac{2}{3}g$. We applied  Nuorium Optimizer Version 21 \cite {nuorium}, the advanced version of NUOPT \cite{nuopt} for these arguments.
  Generally speaking, a smaller value of the least squares gives a better model, 
 as we can imagine from the likelihood function in statistics.


We may have errors caused by our 
assumptions. So, 
 although the computations are 
carried out 
with enough precision,  we can not tell the number of significant digits. The logistic growth is statistically natural in a global view over time. Locally the
boundaries strongly affect our computation and strange absolute dating as we see for the seniority problem among siblings \ref{siblings}. 
The birth year of Te\b{h}ip-tilla is 22.50 ($g=22.5$) years earlier than his sons Enna-mati and \v{S}urki-tilla. 
The  optimum solution of the least squares method with linear inequality constraints
often exists on the boundaries of the region, satisfying the 39783 constraints. The precision of our computation 
works very well  to study the relative chronology;
the seniority among siblings, shown in Section \ref{siblings} and Section \ref{dating} and chronological order of documents and  Section \ref{dating}. 

\subsection{Generation length    for adjustment}\label{scaling}
We can obtain the numerical values for the generation length $\alpha  g$ and with 
the life spans $\alpha U_i$, $i=1,...,10253$, with any positive $\alpha$,  from the numerical values for the generation length $g$
 with life spans $ U_i$, $i=1,...,10253$.

Consider  the generation length $\alpha g$ and the simulated life span  
$\alpha U_i$,  $i=1,...,10253$, for any positive $0<\alpha$.  The assumptions for $\alpha g$, (I), (II), 
and (III) are given  by 
\begin{equation}\label{ls22}
  \left\{
   \begin{array}{ccl}
(b_i  -P_{k_0})&\geq & (b_{f_i}-  P_{k_0}) +\alpha g\\
(b_i  -P_{k_0})&\geq & (b_{m_i}-  P_{k_0}) +\alpha g\\
(d_i - P_{k_0})&\geq&   (P_k -  P_{k_0})\geq ( b_i - P_{k_0})+\frac{2}{3}\alpha g. \\ 
   \end{array} \right.
   \end{equation} 
 where the person $f_i$ is the father  
of the person $i$ and  $m_i$ is the mother   
of the person $i$. 
 Let $b^0_i,  d^0_i, P^0_k$ attain our  least squares problem Eq. \ref{objective}  with  $U_i$, $i=1,..,1053$,
Eq. \ref{ls22} for the generation length $g$. 
 That is to say,   $b'^0_i=-1355+\alpha(1355+b^0)$,  
$d'^0_i=-1355+\alpha(1355+d^0)$
$P'^0_k=-1355+\alpha(1355+P^0) $,
 attain the least squares for $\alpha$. 
For example, the mean of  estimates for the  birth year of 
Te\b{h}ip-tilla son of Pu\b{h}i-\v{s}enni for $\alpha g=25$ is -1426.71 which is obtained from 
-1419.54 (Fig. \ref{tehipfamily} for $g=22.5$) as 
$-1426.71=\alpha (1355-1419.54)-1355$
for $\alpha=\frac{25}{22.5}$.
.

We do not have data for the statistical inference on $g$ in assumptions (I) and (II) 
and take  $g=22.5$  as a possible value. However, our results are 
seniority among siblings, chronological order, and logistic growth 
, 
which do not depend on the value of $g$.

\section{Computation on seniority among siblings for the Te\b{h}ip-tilla family}\label{siblings}

We  see  in p.151 \cite{mai}, 
``The  establishment of synchronisms  among members of the Te\b{h}ip-tilla family can aid in determining the  order of birth among siblings  that in turn  leads to the possibility of determining the identity of chief heirs" and   ``Purves has suggested that this may indicate that Nuzi scribes listed siblings in order of  seniority." 

\subsection{Seniority among siblings obtained by our method is stable and does not depend on random numbers}\label{seniority}

\begin{figure}[ht]
\includegraphics[scale=0.35]{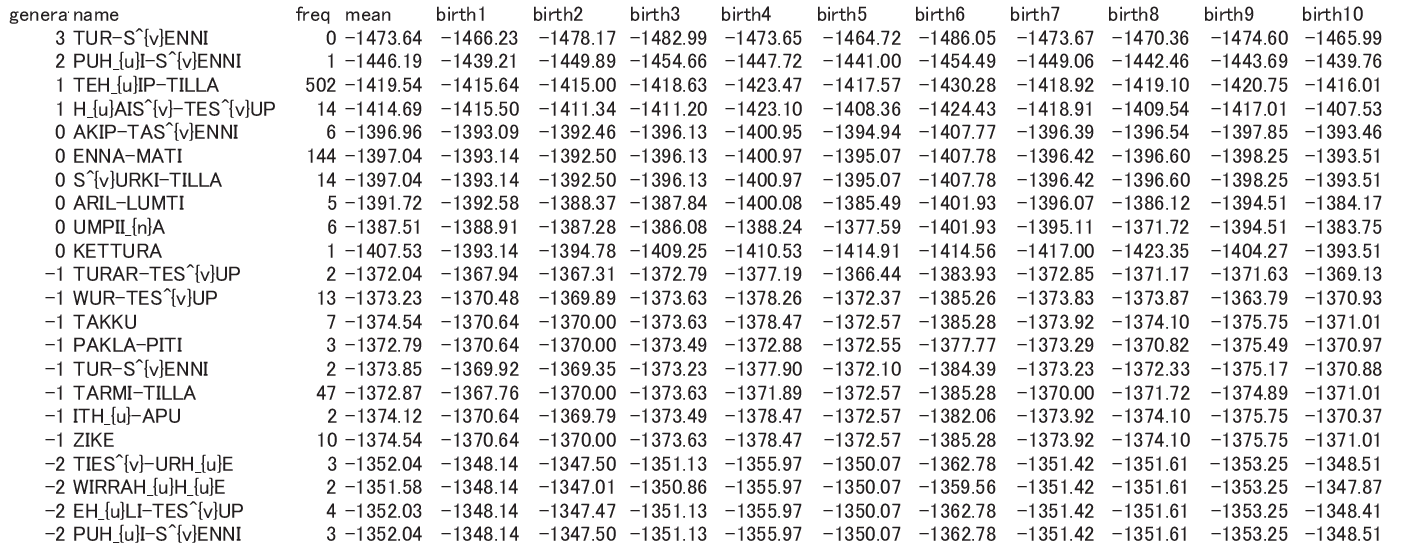}
\caption{Birth years for men of the Te\b{h}ip-tilla family.  For example, we see that Te\b{h}ip-tilla  son of 
Pu\b{h}i-\v{s}enni is  in  502 texts, and  the mean   of birth years by 10 computations, birth1,...,birth10,   is 
-1419.54. }\label{tehipfamily}
\end{figure}

In Fig. \ref{tehipfamily} we give 
$b_i(r)$, for the male person $i$   of Te\b{h}ip-tilla family, for $r$, $r=1,...,10$ 
with the mean. The first  column of numerical values are for the mean $\bar{b}_i=\frac{1}{10}\sum_{r=1}^{10} b_i(r)$. The values for $r$, $r=1,...,10$,  
are given in the columns birth1, ..., birth10.

 Our computational results  
        on the 10 sets of simulated life span in Fig \ref{tehipfamily} do not contradict4c with Maidman's results (p.153 \cite{mai}). 
It seems the constraints strongly work on the computational results.

\begin{enumerate}
\item  `` Te\b{h}ip-tilla was older 
than \b{H}ais-te\v{s}up."

Computational results:
 Let the person $i$ be Te\b{h}ip-tilla  and $j$ be \b{H}ais-te\v{s}up. 
Numerically for all 10  columns $r=1,...,10$ (shown in the columns  birht1,...,bith10), we have 
$b_i(r)\leq b_j(r)$.

\item  ``Te\b{h}ip-tilla's sons were born in the order Enna-mati, 
\v{S}urki-tilla and Akip-ta\v{s}enni."

Computational results:
For the  $i$ of Enna-mati, the $j$ of
\v{S}urki-tilla and the $k$ of Akip-ta\v{s}enni.
Numerically, for  $r=1,...,10$, 
we have $b_i(r)\leq b_j(r)\leq b_k(r)$. 
However $b_i(r)= b_j(r)$, with
$\bar{b}_i= -1397.04$, 
 $\bar{b}_j=-1397.04$, and 
 $\bar{b}_k=-1396.96$.

\item  ``Of Enna-mati's  sons, Takku was probably
the oldest."

Computational results: 
For the  $i$ of Takku, 
the  $j$ of  Tur-\v{s}enni,   
and 
the $k$ of Pakla-piti.
Numerically
we have
$b_i(r)\leq b_j(r), b_k(r)$ , for  $r=1,...,10$. 
We have 
$\bar{b}_i=-1374.54$,
 $\bar{b}_j=-1373.85$, and
$\bar{b}_k=-1372.79$.

Remark. The seniority between 
Tur-\v{s}enni and Pakla-piti is not given in 
p.153 \cite{mai}.  Also in  our computational Fig \ref{tehipfamily} Tur-\v{s}enni 
is older in the columns birth4, birth6, and birth8, while in the remaining  7 columns 
Pakla-piti is older.

\item  ``It seems that of the two sons of \b{H}ais-te\v{s}up, 
 Aril-lumti was older than Umpiya."

Computational results: 
For the  $i$ of Aril-lumti and the  $j$ of Umpiya.
Numerically, we have 
$b_i(r)\leq b_j(r)$,  $r=1,...,10$, 
We have 
 $\bar{b}_i=-1391.72$ and  $\bar{b}_i=-1387.51$. 
\end{enumerate}

\subsection{Computation  for the problems suggested by Maidman}\label{problem}
  We study the following two arguments 
of Maidman (p.154 \cite{mai}).
\vspace{0.2cm}
 
 1.   ``Of the known sons of Akip-ta\v{s}enni, Wur-te\v{s}up would appear to be older than 
Turar-te\v{s}up, if one considers the number of texts identified with each. Yet Wur-te\v{s}up's role is consistently that of either witness for, or ceder of land to, his first cousin Tarmi-tilla. Turar-te\v{s}up, on the other hand, owns land (JEN 294, JENu 546?). 
..."     
\vspace{0.2cm}

Computational results:
For the  $i$ of Wur-te\v{s}up and 
the $j$ of
Turar-te\v{s}up.
Numerically,  
$b_i(r)\leq b_j(r)$,   $r=1,...,10$. We have 
$\bar{b}_i=-1373.23$ and 
$\bar{b}_j=-1372.04$. 
Our computation supports that 
Wur-te\v{s}up is older 
than  
Turar-te\v{s}up.
\vspace{0.2cm}
 
2. ``...The impression one gets, therefore, is that though Zike may have been active in his own right, external real 
estate texts suggest that Tarmi-tilla was more active and hence more active and hence probably older. JEN 538:6-8 in fact introduces the brothers in the order Tarmi-tilla, Zike. Yet JEN 662:1 and HSS 13221:66-67 use the order Zike, Tarmi-tilla." 
\vspace{0.2cm}

Computational results: 
For the  $i$ of  Zike and the $j$ of Tarmi-tilla. 
Numerically,
, we have
$b_i(r)\leq b_j(r)$,   $r=1,...,10$. We have 
$\bar{b}_i=-1374.54$ and 
$\bar{b}_j=-1372.87$.  Our computation supports that Zike is older than Tarmi-tilla.

\section{Dating for Assyriological  results of Maidman}\label{dating}
 
\subsection {Distribution of the published years  on Enna-mati son of Te\b{h}ip-tilla}\label{distribution}
\begin{figure}[ht]
\includegraphics[scale=0.70]{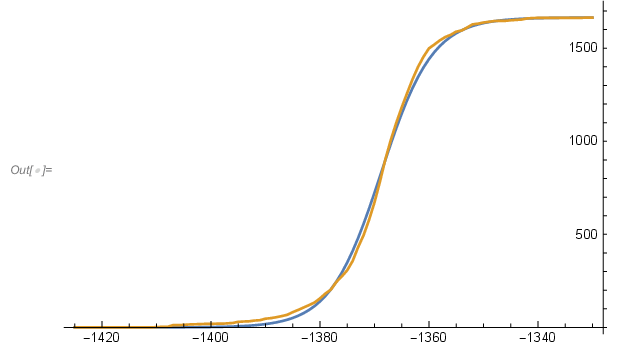}
\caption{ Logistic growth of the number of published texts each year obtained from RNPN2.  
The maximum log-likelihood value for the logistic 
distribution is -6130.72,  while that for the normal distribution is -6220.25.
The maximum log-likelihood for normal distribution is attained by the parameters $\mu=-1366.94, \sigma=10.1447$, while those for the logistic distribution are  $\mu=-1366.84, \beta=5.2842$.
The colored line shows the data.   The black line shows the estimated logistic distribution. 
}\label{logistic}
\end{figure}

\begin{figure}[ht]
\includegraphics[scale=0.70]{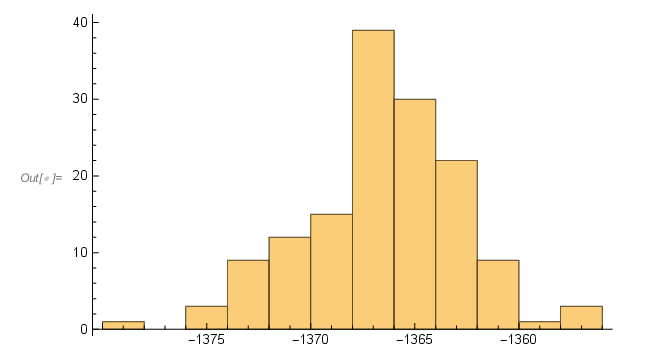}
\caption{Histogram  for published years of 144 texts for Enna-mati  son of 
Te\b{h}ip-tilla obtained from RNPN2.     Maximum log-likelihood for
normal distribution -391.705 by $\mu=-1366.5, \sigma=3.67383$.
Maximum log-likelihood for
logistic distribution is -390.218 by $\mu=-1366.36, \beta=2.03066$.
}\label{histogramenna}
\end{figure}

 The number of tablets seems to have increased by  
logistic growth until saturation (\cite{u, umit}) for RNPN2, as in our previous study  \cite{umit} for NPN and seems to show 
 nearly at the saturation stage of logistic growth, the town of Nuzi was destroyed by  Assyria (Fig \ref{logistic}). 
We show  the distribution of the published years for
 Enna-mati  son of Te\b{h}ip-tilla (Fig \ref{histogramenna}),  which will help to see the detail of the logistic growth.
 Most of the 1665 documents are published in the period of two generations,  as we can see from that Te\b{h}ip-tilla  son of Pu\b{h}i-\v{s}enni is listed in 502 documents and    Enna-mati son of Te\b{h}ip-tilla  is listed in 144 documents (Fig  \ref{tehipfamily}).

\begin{figure}[ht]
\includegraphics[scale=0.33]{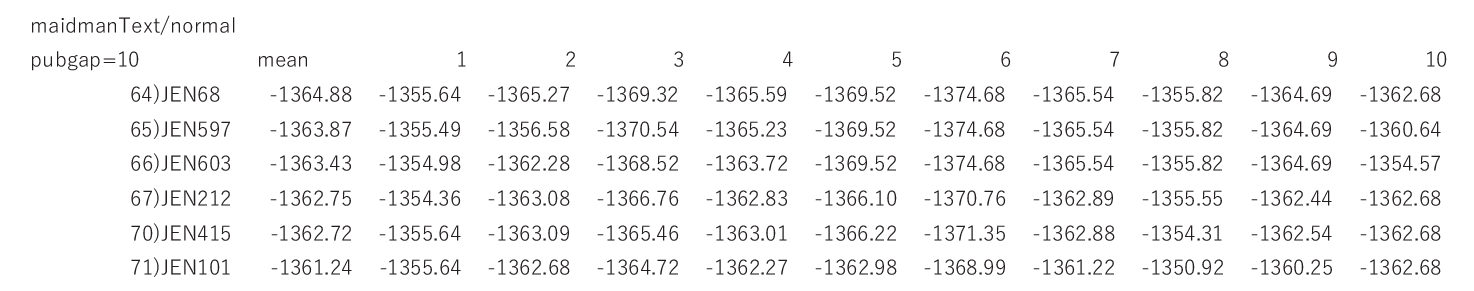}
\caption{ The published years of the 6 texts in Chapter Four, Decline and the Fall of a Nuzi Family, \cite{mai3}. 
$P_k(r)$ for the text number $k$   of each of the above 6 texts,
 for $r$, $r=1,...,10$ are given 
with the mean in the first
  columns.   }
\label{hismeya}
\end{figure}

\subsection{ Chronological order on "The decline and fall of a Nuzi family":\cite{mai3} }\label{hismeyasec}
Here, we study a series of texts  for 
``The stronger families at times exploit certain other families repeatedly,  
 systematically, and extensively"  (p.9, Maidman \cite{mai3}). Many texts, including the following texts, may accumulate to make the logistic growth of the properties or lands of powerful families. 
The 6 texts for land transactions,
 \#64 (JEN68),
\#65 (JEN597),
\#66 (JEN603),
\#67 (JEN212),
\#70 (JEN415),
\#71 (JEN101),
 out of 9 texts for the series (\#63-\#71) Chapter Four, Decline and the Fall of a Nuzi Family\cite{mai3}),  are in NPN. 
In p.143 \cite{mai3}, ``Thus the text \#63 should represent an early stage in the decline  of the family of 
\b{H}i\v{s}meya son of It\b{h}i\v{s}ta."	
 The text \# 63. HSS XIII, 62 (SMN 62), given  p.146 \cite{mai3}) as 		 					
"(1-8) 3 homers of barley belonging to Enna-mati were delivered on interest to \b{H}i\v{s}meya son of It\b{h}i\v{s}ta.
\b{H}i\v{s}meya shall return the barley at harvest time, interest included, to Enna-mati."
is not listed in NPN, and we can not estimate the published year. 
Let us study some parts of 
\# 64.JEN 1, 68 (JENu 622) (p. 148 \cite{mai3},											              

 "(1-5) Tablet of adoption of \b{H}i\v{s}meya son of It\b{h}i\v{s}ta and of Ussen-naya daughter of Enna-milki.	They adopted Enna-mati son of Tehip-tilla.

(6-12) \b{H}i\v{s}meya and Ussen-naya gave to Enna-mati as his inheritance share a 1.6 homer field in the town of Tente, to the east of the road to the town of Natmane, to the north of the field of [SamasRI?], 									and to the west ... of the enclosure.											            

(13-16) And Enna-mati gave to \b{H}i\v{s}meya and to Ussen-naya as their gift 1 garment, 1 leather ziyanatu-blanket, and 5[+n?] sheep.										              
(21-23) Whoever amongst them abrogates (this contract) shall pay 2 minas of silver (and) [2] minas of gold.	"									             

 Let us study some parts of the \#71 ( p. 161 
\cite{mai3}. 
					
"(1-14) Thus \v{S}arra-\v{s}ad\^{u}ni son of It\b{h}i\v{s}ta: ``I shall not raise a claim against them, to wit whatever awiru-land, whatever (empty?) 
building plot and built-up plot (lit. "a plot of built structures") in the midst of the town of Tente, 
(whatever) threshing floor, orchard, and halahwu-plot in the hinterland, in the town(ship) of Tente, 										
which \b{H}i\v{s}meya and Akiya, my brothers, have given to Enna-mati son of Te\b{h}ip-tilla.

(15-18) Should \v{S}arra-\v{s}ad\^{u}ni raise a claim against land described in 
(lit. `according to the mouth of') this  tablet, 										
he shall pay to Enna-mati 1 mina of silver (and) 1 mina of gold.

(19-20) And Enna-mati has given to \v{S}arra-\v{s}ad\^{u}ni as his 
(i.e., as \v{S}arra-\v{s}ad\^{u}ni's gift 1(?) fine 
  sheep(?)."	
							
(21-30) Before Zunzu son of Initya;  
before Apu\v{s}ka son of Ithip-\v{s}arri; ...
."


In p. 144 \cite{mai3}, "Probably, the earliest of these are texts \# 64, \#65 and \#66, not necessarily in the order." (p.143 i\cite{mai3}).  
"text \#70 must be the last or next to the last", 
``The following document 
\#71 is most probably the last tablet of this series." 
An     analogous observation to  Section \ref{siblings}  for Fig \ref {hismeya} gives the chronological order \#64 (JEN68),
\#70 (JEN415),
\#71 (JEN101). 
 Our computation (not shown here) supports,   "It\b{h}i\v{s}ta's first-born son (probably ), 
\b{H}i\v{s}meya,(p. 143 \cite{mai3}."

Our estimated ages of   
{\small \verb|H_{u}IS^{v}MEI_{n}A|},	
{\small \verb|ENNA-MATI|} and 
{\small \verb|ZUNZU|}  (scribe), at JEN  101, are  19.9, 35.8 and 40.4, respectively,  
which  makes us imagine the scene of the contract of 
3500 years ago.

\section{Discussion and conclusion}\label{discussion} 

We can apply the argument in Section \ref{scaling} if we have a more reasonable  value other than $g=22.5$. 
Our problems  require  precision which cannot be accomplished  by 
radiocarbon dating.  
Here we make RNPN2\cite{u2} adding kinships (Section \ref{adding}) to eNPN.   The representation of family trees to compute the sequential unifying is shown in \cite{umi,u} with examples. To make RNPN2 from eNPN, we should solve many computational problems one by one.  
To release  computer programs \cite{u2}  to make RNPN2 from eNPN  is our next project. 

The NPN published in 1943 should be updated. Adding the kinships of two family trees, the Te\b{h}ip-tilla family tree and the Kizzuk family tree \cite{mai3}, to reflect the studies after 1943 greatly advances our computational study toward understanding the Assyriological results of Maidman \cite{mai,mai3}.
The updated NPN should be electronic. Our study could be a base for updating the NPN 
and shows a method to add information to eNPN 
  one by one (Section \ref{adding} Section \ref{introduction}).
Much more data have been accumulated since the publication of NPN in 1943, for example, \cite{cg}. 
 The  
Te\b{h}ip-tilla family tree constructed just based on NPN  is with 15 mpersons \cite{mak,umi,umit}.  
The updated  NPN    will reconstruct the Te\b{h}ip-tilla  family tree of 36 
persons 
\cite{mai, mai3}.


NPN is carefully made, well-designed, and surprisingly useful  
for our study.   We started the present study in 1995 by using a supercomputer \cite{iium}. 
Now the computer
hardware and software advancement enables us to complete the same task on a personal computer.
Artificial intelligence will help to update NPN to include the Nuzi studies after  1943.
Artificial intelligence will be able to learn to translate the written
language of ancient civilizations \cite{mi} and to learn 
 to make the name index as NPN for other archives.
Our method may be applied   
to archives listed in Chapter ``The clay tablet Archives"  \cite{po}, 
once we have the name index in the format of  NPN.

 {\bf Acknowledgments} The authors  thank   Maynard  P. Maidman for his kind introduction to the Assyriological results,  helpful comments, suggestions, and discussions. The authors  thank  Takahito Tanabe, an author  of 
 \cite {nuopt, nuorium, yyt},  for his   suggestions on Section\ref{siblings} and thank his computational results using Nuorium Optimizer \cite{nuorium} in  Section \ref{least}.
 We  thank Shuhei Mano and Howell tong for their careful reading. 
  
  Y. I. thanks   Joel E Cohen for the annual visits and discussions at Rockefeller University and for sending  Matthew W. Stolper's helpful comments.
 This work is in part supported by 
 JSPS Kakenhi  07207245, JSPS Kakenhi
   09204245,   JSPS Kakenhi 23540177 and 
   ISM Cooperative Research Project (2011- ISMCRP -1022). 
 Y.I. is supported in part by
US National Science Foundation Grant DMS 0443803 to Rockefeller University.


\begin{thebibliography}{}
\bibitem{a} 
M. J. Aitken, Science-based dating in archaeology. Routledge, 2014.

\bibitem{cg}
E. Cassin and J. Glassner, Anthroponymie et Anthropologie de Nuzi, vol. 1: Les Anthroponymes (1977).


\bibitem{ch}
E. Chiera,     Joint expedition with the Iraq Museum at Nuzi, 
American Schools of Oriental Research; Publications of the Baghdad School Texts,  Vol. II, 
 (P. Geuthner, Paris, 1930). 



\bibitem{dd}
G. Dosch,  and K. Deller,  Die Familie Kizzuk, Sieben Kassitengenerationnen in Tertena und Suriniwe, SCCNH 1 (1981) 91-113.







\bibitem{gpm}
 I. J.Gelb,  P. M. Purves, and A. A. MacRae,  Nuzi 
Personal names, The University of Chicago Press,  Chicago Illinois, 1943, 
http://oi.uchicago.edu/pdf/oip57.pdf.

\bibitem{hzg}
H.Han, H. Zha,  and  C. L. Giles,   Name disambiguation in author citations using a 
k-way spectral clustering method. In Digital Libraries, 2005. JCDL'05. 
Proceedings of the 5th ACM/IEEE-CS Joint Conference on.  IEEE, 2005 pp. 334-343.


\bibitem{iium}
 Y. Itoh, M.  Ishiguro, S. Ueda, and K.  Makino,  
 Estimating population from the kinship 
data in an ancient society  
(CDROM),  Report of Research Grant 09204245, 
 Ministry of Education, Science and Culture of 
  Japan, 1998 (in Japanese).
 
\bibitem{la}
E. R. Lacheman, Nuzi Personal Names: Review Article. 
Journal of Near Eastern Studies
 8 ( 1949),  48-55. 





\bibitem{laa}
W. F. Libby, E. C. Anderson,  and  J. R. Arnold,  Age determination by radiocarbon content: world-wide assay
of natural radiocarbon, Science, 109 (2827),(1949) 227-8.


\bibitem{mai}
M. P.  Maidman, The Te\b{h}ip-tilla  family of Nuzi -a genealogical reconstruction, 
Journal of Cuneiform Studies 28 (3) (1976) 127-155 .

\bibitem{mai2}M. P. Maidman, 
Correction ``The Te\b{h}ip-tilla Family of Nuzi: A Genealogical Reconstruction,"
 Journal of Cuneiform Studies  29 (1977) 64.
 
\bibitem{mai3}
M. P.  Maidman, Nuzi texts and their uses as  historical evidence, 
 Society of Biblical Literature, Atlanta,  2010.

\bibitem{mai4}
M. P.  Maidman,   Nuzi, the Club of the Great Powers, and the Chronology of the Fourteenth Century, KASKAL 8 ( 2011) 77-139.

\bibitem{mai5}M. P.  Maidman, Personal communication, (August 2020).
\bibitem{mai6}M. P.  Maidman, Personal communication, (February  2021).
\bibitem{mak}
 K.   Makino,  Social change reflected in the source of false adoption contracts 
at Nuzi.  Shigaku 60(1)  (1991) 91-119 (in Japanese with English title). 

\bibitem{mi}
R. Mitchum,
How AI could help translate the written
language of ancient civilizations, uchicago news,  March 11, 2020.



\bibitem{mo}
 M. A.    Morrison, and  D. I. Owen,  Studies on the Civilization and Culture of Nuzi 
and the Hurrians in Honor 
of Ernest R. Lacheman on his Seventy-First Birthday, April 29, 1981, Eisenbrauns, 
 Winona Lake, Indiana, 1981.  

\bibitem{nuopt}
 NUOPT, version 9,   Mathematical Systems Inc. 2007.
\bibitem{nuorium}  Nuorium Optimizer, version 21,  NTT DATA  Mathematical Systems Inc. 2019.
\bibitem{po}
E. Posner,  Archives in the ancient world. Harvard University Press, 2013. 

\bibitem{rd}
Ramsey, C. Bronk, Michael W. Dee, Joanne M. Rowland, Thomas FG Higham, Stephen A. Harris, Fiona Brock, Anita Quiles, Eva M. Wild, Ezra S. Marcus, and Andrew J. Shortland, Radiocarbon-based chronology for dynastic Egypt,  Science 328, no. 5985 (2010) 1554-1557.
\bibitem{sm}
B. Scherer, R. D. and Martin,  Modern Portfolio Optimization with NuOPT, S-PLUS, and S+ Bayes,  Springer Science \& Business Media, 2005.
\bibitem{starr}
R.  F. S.  Starr,  
Nuzi;  report on the excavation at Yorgan Tepa near Kirkuk, Iraq, conducted 
by Harvard University in conjunction with the American Schools of Oriental Research 
and the University Museum of Philadelphia, 1927-1931,  Volume 2,  Plates and Plans, 
Harvard University Press (1937).  

\bibitem{st}
M. W. Stolper. Entrepreneurs and Empire: The Mura\v{s}\^{u} Archive, the  Mura\v{s}\^{u}Firm, and Persian Rule in Babylonia. Nederlands historisch-archaeologisch instituut, 1985.

\bibitem{tfg}
G. Tilahun, A. Feuerverger, \& M. Gervers, Dating medieval English charters. The Annals of applied statistics, 6(4) (2012)  1615-1640.

\bibitem{tsuc} T.  Tsuchiya,  
Interior-point Algorithms,  Information Geometry
and Optimization Modeling,  
Proceedings of the Institute of Statistical Mathematics  61  (2013) 3-16 (in Japanese with English Summary).


\bibitem{u}  S. Ueda, Statistical Mathematics Approach to Human Sciences,  
(PhD Thesis), Graduate University for Advanced Studies, 2010  (in Japanese). 




\bibitem{u2} S. Ueda, Computation on    Nuzi Personal Names in Ancient Mesopotamia (ready to be uploaded to a repository).

 \bibitem{umi}
S. Ueda, K.  Makino, and Y. Itoh,   Reconstructing family trees in ancient population 
from the Nuzi personal names, 
Proceedings of the Institute of Statistical  Mathematics 53 (2005) 285-295 
(in Japanese with English summary). 


\bibitem{umit}
S. Ueda, K. Makino, Y. Itoh,  \& T. Tsuchiya,  Logistic growth for the Nuzi cuneiform tablets: Analyzing family networks in ancient Mesopotamia. Physica A: Statistical Mechanics and its Applications 421 (2015) 223-232.


\bibitem{yyt}
H. Yamashita, H. Yabe \& T. Tanabe,  A globally and superlinearly convergent primal-dual interior point trust region method for large scale constrained optimization. Mathematical  Programing  102 (2005) 111-151 .

\bibitem{wlk}
A. Wagner, Y. Levavi, S. Kedar, K. Abraham, Y. Cohen, \& R. Zadok.  Quantitative Social Network Analysis (SNA) and the Study of Cuneiform Archives: A Test-case based on the  Mura\v{s}\^{u} Archive. Akkadica, 134(2) (2013) 117-134.





\end{thebibliography}
\end{document}